\documentclass[10 pt, letterpaper]{article}

\usepackage{amssymb}
\usepackage{amsmath}
\usepackage{graphicx}
\usepackage{times}
\usepackage[T1]{fontenc}

\newcommand\highertop{\rule{0pt}{3.1ex}}

\begin{document}
\date{}

\title{Design of a small-scale prototype for research in\\ airborne wind energy
\thanks{This manuscript is a preprint of a paper submitted for possible publication on the IEEE/ASME Transactions on Mechatronics and is subject to IEEE Copyright. If accepted, the copy of
record will be available at \textbf{IEEEXplore} library: http://ieeexplore.ieee.org/.}
\thanks{This research has received funding from the California Energy Commission under the EISG grant n. 56983A/10-15 ``Autonomous flexible wings for high-altitude wind energy generation'', and from the European Union Seventh Framework Programme (FP7/2007-2013) under grant agreement n. PIOF-GA-2009-252284 - Marie Curie project ``Innovative Control, Identification and Estimation Methodologies for Sustainable Energy Technologies''. The authors acknowledge SpeedGoat$^\circledR$'s Greengoat program.\newline
Corresponding author: Lorenzo Fagiano, fagiano@control.ee.ethz.ch.}}

\author{Lorenzo~Fagiano\thanks{L. Fagiano is with the Automatic Control Laboratory, Swiss Federal Institute of Technology,Zurich, Switzerland. E-mail: fagiano@control.ee.ethz.ch.}$\,$ and  Trevor~Marks\thanks{T. Marks is with the Dept. of Mechanical Engineering, University of California at Santa Barbara, USA. E-mail: tmarks@engineering.ucsb.edu.}
}
\maketitle

%

\begin{abstract}
Airborne wind energy is a new renewable technology that promises to deliver electricity at low costs and in large quantities. Despite the steadily growing interest in this field, very limited results with real-world data have been reported so far, due to the difficulty faced by researchers when realizing an experimental setup. Indeed airborne wind energy prototypes are mechatronic devices involving many multidisciplinary aspects, for which there are currently no established design guidelines. With the aim of making research in airborne wind energy accessible to a larger number of researchers, this work provides such guidelines for a small-scale prototype. The considered system has no energy generation capabilities, but it can be realized at low costs, used with little restrictions and it allows one to test many aspects of the technology, from sensors to actuators to wing design and  materials. In addition to the guidelines, the paper provides the details of the design and costs of an experimental setup realized at the University of California, Santa Barbara, and successfully used to develop and test sensor fusion and automatic control solutions.
\end{abstract}


\section{Introduction} \label{S:intro}

In the last decade, an increasing number of research groups and companies worldwide have been developing a new concept of wind energy generation, named airborne wind energy (AWE), see e.g. \cite{Makani,skysails,ampyx,windlift,kitenergy,CaFM07,IlHD07,CaFM09c,TBSO11,BaOc12}, as well as \cite{FaMi12} for an overview. Airborne wind energy systems aim at harnessing the wind blowing up to 1000 m above the ground, using tethered wings flying fast in crosswind conditions, i.e. roughly perpendicular to the wind flow. This recent interest in airborne wind energy is fostered by a series of factors, both technical and socio-political. On the technical side, the development of advanced solutions in fields like materials, mechatronics, and power conversion have made these concepts, which firstly appeared in the late 1970s \cite{Mana76,FlRo79,Loyd80}, technically feasible today. On the socio-political and economical side, the research and development of novel forms of renewable energy is driven by the actual energy situation and environmental issues, caused by the extensive use of fossil fuels, that represent one of the most urgent challenges on a global scale but also an important market opportunity for renewable technologies.

Despite the mentioned recent developments, several technical aspects of airborne wind energy still require research and development, in order to definitely assess the viability of this concept and to transform it into an industrial product. Tether technology, aerodynamics and wing design, sensors, control and energy conversion are all fields where research activities specifically aimed at airborne wind energy are needed, either to solve technical bottlenecks or to improve off-the-shelf solutions that are being already used. However, it is not easy to carry out research activities in this field, due to at least two interrelated aspects. First, theoretical and numerical studies have now reached a quite mature stage  \cite{IlHD07,CaFM09c,TBSO11,BaOc12,FaMP09,FaMP11}, so that new results appear to be difficult to obtain without experimental tests. Second, it is not trivial to carry out experimental tests with an airborne wind energy system, due to the difficulty to obtain a prototype to be used for testing. Differently from well-established renewable energy technologies like solar or conventional wind, there is no airborne wind energy system that can be ordered to carry out experiments, as well as no testing facilities that can be used. Indeed several operating prototypes have been built by the mentioned research groups and companies (see e.g. \cite{Makani,skysails,ampyx,windlift,kitenergy,enerkite}), however they are  generally not accessible and their design and functioning is not available for replication. Airborne wind energy systems are mechatronic devices encompassing aspects from different disciplines, 
such that researchers with specific competencies in any one of these areas have to overcome non-trivial barriers, before being able to apply their core expertise and carry out experiments. The scientific literature lacks contributions detailing the design guidelines, construction costs, and basic operation of these systems, hence making research in airborne wind inaccessible, unless a consistent amount of time and money is dedicated to overcome the initial barriers. 

In this paper, we aim to partially fill this gap by providing the guidelines and the details of the complete design of a small-scale prototype to study airborne wind energy systems, including power supply, mechanics, actuators, low-level control systems, lines and wings, sensors and human-machine interface. Albeit the described system is not able to actually generate electricity, its low cost, transportability and easiness of use make it a good testing system for many aspects, like wing design and aerodynamics, line wear, sensor fusion, and control design. Moreover,  we describe the basic maneuvers that can be carried out with the described system, using a manual control of the wing through a human-machine interface. Finally, we comment on how the proposed design can be complemented by adding energy conversion capabilities. The aim is to provide researchers, who have basic knowledge of mechanical and/or electrical engineering, with a  reference to realize a significant testing system in a short time and with limited funds. We present both general design guidelines for a prototype and the specific solution realized at the University of California, Santa Barbara, in the first six months of a one-year-long research project. Such a prototype has been used successfully to develop sensor fusion algorithms and a feedback controller able to achieve consistently autonomous flight paths, see \cite{FHBK12_arxiv,FZMK13_arxiv}, as well as  \cite{Wing_movie} for a short movie of the experimental tests.

The paper is organized as follows. In section \ref{S:basics} we provide a concise description of the considered concept of airborne wind generators, and in section \ref{S:layout} we present the layout and the main design guidelines of the prototype. In section \ref{S:design} we give the details of all of the aspects of the proposed design, including a cost breakdown, and we describe the basic operation of the system. We present experimental results throughout the paper to support the design considerations. We draw conclusions and comment on future developments in section \ref{S:conclusion}.

\section{Basics of airborne wind energy}\label{S:basics}

\begin{figure}[!hbt]\centering{
\includegraphics[width=6cm]{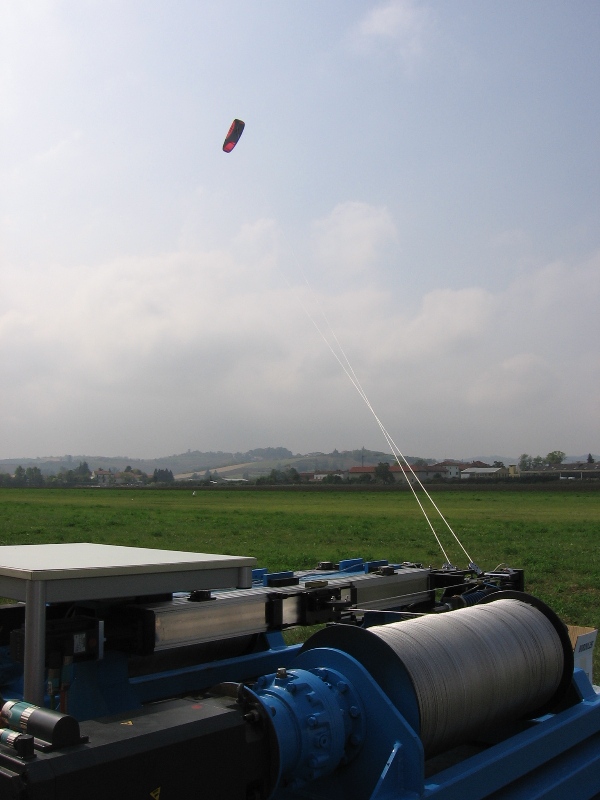}}
\caption{AWE small scale prototype operating near Torino,
Italy.}\label{F:proto_1}
\end{figure}

Airborne wind energy (AWE) systems aim at producing wind energy using a wing or an aircraft linked to the ground by tethers, hence reducing the construction costs with respect to conventional wind turbines and making high-altitude winds (up to 1000 m above the ground) reachable. The lift force generated by the aircraft immersed in the wind is sufficient to keep the system airborne and to generate energy, through one of different possible mechanisms. Several technologies based on this concept are being developed worldwide, with different approaches depending on the type of aircraft and on whether mechanical power is converted into electricity on the ground or onboard, see \cite{FaMi12} for an overview. In this work, we focus on AWE systems with ground-level generators, which exploit the aerodynamic lift generated by a wing, either rigid or flexible, linked to the ground by three lines. In these systems, the wing's lines are wound around one or more winches installed on the ground and linked to electric generators. Energy is obtained by continuously performing a two-phase cycle, composed by a \emph{traction phase}, during which the lines are unrolled under high traction forces and the generators, driven by the rotation of the winches, produce electricity, and by a subsequent \emph{passive phase}, when the electric generators act as motors, spending a fraction of the previously generated energy to recoil the lines. The traction phase has to be carried out in the so-called ``crosswind'' conditions, i.e. with flight trajectories that are roughly perpendicular to the wind flow. The passive phase can be carried out in different possible ways, depending on whether the wing's pitch can be changed or not, see \cite{FaMi12} for more details. Fig. \ref{F:proto_1} shows a small-scale prototype of one such system, built at Politecnico di Torino, Italy.

AWE generators rely heavily on automatic control for their operation. The control system collects the information coming from a series of sensors in order to compute suitable control inputs, which are sent to the available actuators in order to achieve the desired control objectives. The first of such objectives is to keep the wing airborne, by stabilizing its open-loop unstable flying paths. The second objective is to realize the mentioned energy generation cycle, by maximizing the produced average power. Among the possible actuation solutions, we focus here on ground-based ones, which manipulate the length of the wing's lines at ground level in order to influence the flight trajectory. In particular, in three-lined systems the two outermost lines, or ``steering lines'', are used to steer the wing: a longer right line with respect to the left one gives rise to a counter-clockwise turn of the wing (as seen from the ground), and vice-versa. The center line (``power line'') sustains most of the generated force and its length can be adjusted to influence the wing's pitch angle, in order to power or de-power the wing.

In the next sections, we provide the main guidelines to design a small-scale prototype of an airborne wind energy system, which can be used to study several aspects of this technology. Moreover, in section \ref{S:design} we describe in details a prototype designed and built at the University of California, Santa Barbara, which has been recently used to study sensor fusion and automatic control strategies. 
In order to have a low-cost experimental setup and to be able to test without particular flight permissions in most areas, we focus our attention on a system without energy generation capabilities (i.e. fixed length of the lines) and with up to 50$\,$m of line length. We provide in the last section of the paper some guidelines on how this setup can be expanded to include also line reeling and generators. Moreover, we focus on a prototype setup to be used for research and development activities, with a high versatility which allows one to easily modify the system (for example by changing/adding sensors or actuators or by modifying the mechanical frame), rather than on a definitive layout.

\section{System layout and design guidelines}\label{S:layout}

\begin{figure*}[!hbt]
\centering{
\includegraphics[width=14cm]{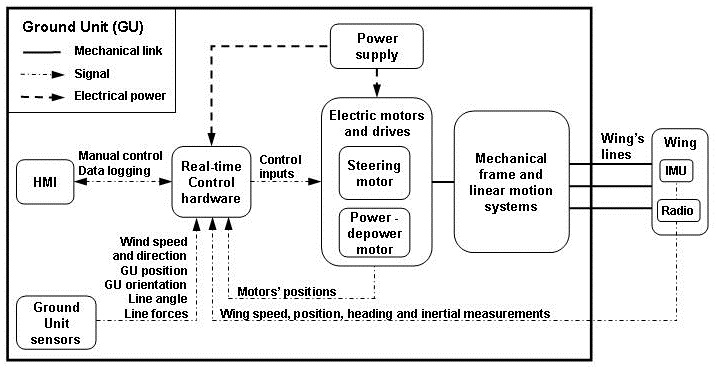}}
\caption{Layout of the prototype. The low-power supply links for the sensors are not depicted in the diagram.}\label{F:layout}
\end{figure*}

The layout of the considered prototype system is depicted in Fig. \ref{F:layout}. For simplicity, we denote as ground unit (GU) all the parts placed on the ground. The  functional subsystems of the prototype are the mechanical frame, the linear motion systems (LMSs) used to actuate the control inputs, the electric motors and their drives, the power supply system, the real-time control hardware, the human-machine interface, the wing and its lines, finally the ground and onboard sensors and radio link. We will now provide the main design guidelines for each of these systems.
\subsection{Mechanical frame and linear motion systems}\label{SS:frame}
\begin{figure}[!hbt]
\centering{
\includegraphics[width=5cm]{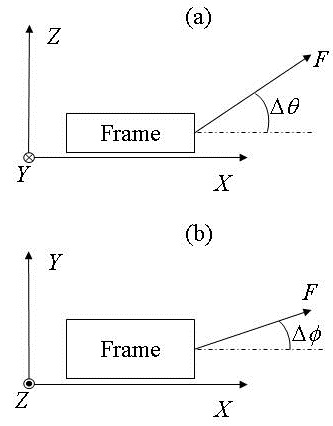}}
\caption{Scheme of the total load $F$ exerted by the wing's lines on the frame during crosswind flight. (a) side view of the frame; (b) top view. The frame $X$ axis is assumed to be aligned with the wind velocity vector.}\label{F:load_config}
\end{figure}

\begin{figure}[!hbt]
\centering{
\includegraphics[width=5cm]{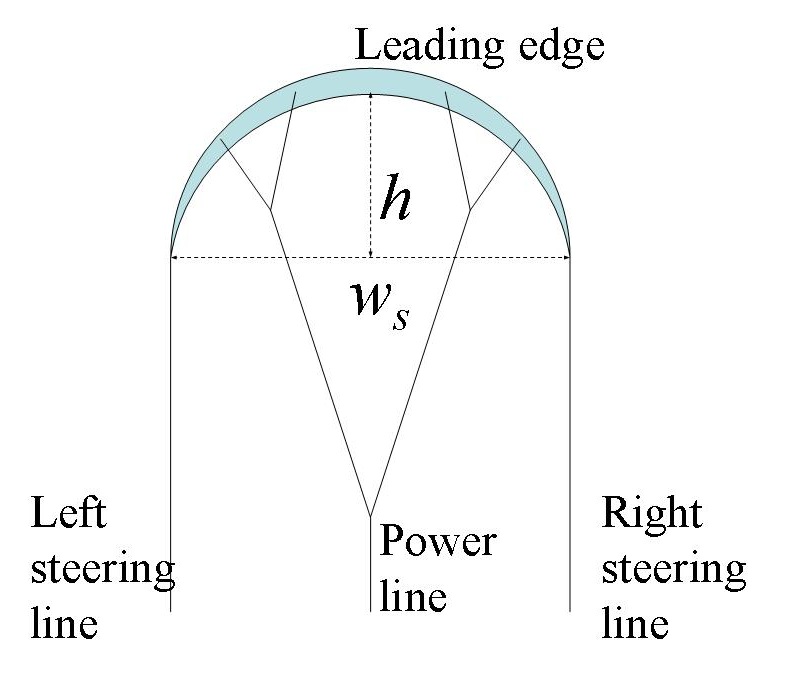}}
\caption{Scheme of the front view of a three-lined curved flexible wing, showing the wingspan $w_s$ and wing's height $h$.}\label{F:wing_geom}
\end{figure}

The functions of the mechanical frame are to withstand and transfer to the prototype base (usually a light truck or a trailer, for easy transportability) the loads exerted by the wing's lines, and to provide mounting points for the linear motion systems and the electric motors, as well as for several ground sensors such as  load cells, line angle sensor,  anemometer and  ground GPS. As regards the loads applied by the wing's lines, we give next the guidelines to compute, for a given wing, the total generated force as a function of the wind speed and its partitioning among the three lines. The flight condition that yields the highest forces is a fast crosswind motion of the wing. In such a condition, the wing's path is typically a cyclic figure-eight trajectory that gives rise to periodic loads acting on the GU frame. The loads have periodic nature both in magnitude and direction. A scheme of the load configuration is depicted in Fig. \ref{F:load_config},
where the longitudinal axis $X$ of the GU is assumed to be parallel to the ground and aligned with the absolute wind velocity vector. In this situation, the maximum lateral angle of the lines with respect to the $X$ axis, $\Delta\phi$, can be assumed to be equal to $\pm\pi/4\,$rad for the purpose of dimensioning of the frame, while the elevation range $\Delta\theta$  (i.e. the maximum angle that the lines can have relative to the ground) can be taken as $1\,$rad. The traction force decreases dramatically beyond such angular ranges, so that lines' directions outside the described ranges are not representatives of the most critical loading conditions on the frame. For a wing with effective area $A$, lift coefficient $C_L$, and aerodynamic efficiency $E$, the peak total line force $\overline{F}$ to be expected as a function of the wind speed $W$ is given by (see e.g. \cite{FaMP11} and the references therein):
\begin{equation}\label{E:total_force}
\overline{F}=\frac{1}{2}\,\rho\,A\,C_L\,E^2\,\left(1+\dfrac{1}{E^2}\right)^{\frac{3}{2}}\,W^2,
\end{equation}
where $\rho\simeq1.2\,$kg/m$^3$ is the air density. Since the considered system's layout does not include winches and generators, the wing's lines are attached directly to the frame. The force value given by equation \eqref{E:total_force} can then be used to properly dimension the attachment points and the frame components, according to well-established mechanical engineering practices \cite{Juvinall11}, considering the intended operating conditions in terms of wind speed and the characteristics of the wings that will be employed. A safety factor of $2$ on the peak force should be used in the design to account also for the effects of inertia and wind gusts, which can increase the peak force. The lowest total force during a figure-eight path can be taken as $\underline{F}=\frac{1}{4}\overline{F}$. For a fixed size of the wing, the period of the oscillations between $\overline{F}$ and $\underline{F}$, denoted as $T_L$, depends on the aerodynamic efficiency and on the wind speed:  in general the higher these values, the larger the wing's speed and the smaller the value of $T_L$ for a fixed length of the path:
\begin{equation}\label{E:force_oscillations}
T_L=\dfrac{1}{2}\,\dfrac{L}{E\,W},
\end{equation}
where $L$ is the length of the flown path in meters. The latter depends on several factors, including the size of the wing (in particular its span, $w_s$, depicted in Fig. \ref{F:wing_geom}) and the chosen control algorithm (which influences the shape of the flown paths).
The coefficient $\frac{1}{2}$ in \eqref{E:force_oscillations} is due to the fact that the frequency of oscillation of the force is twice that of a single figure-eight trajectory, since the maximal and minimal force values are hit twice in a full path. To find a lower bound on $T_L$ for a given wing, one can consider the minimum turning radius $\underline{r}$:
\begin{equation}\label{E:min_radius}
\underline{r}\simeq2.5\,w_s,
\end{equation}
and then approximate the smallest figure-eight path that can be flown as the sum of the circumferences of two complete loops with minimum radius:
\begin{equation}\label{E:path_length}
L\simeq2\left(2\pi\underline{r}\right).
\end{equation}
Finally, the partitioning of the total force among the three lines depends on the wing's design and bridling configuration; typically the power line accounts for 55\% to 75\% of the total load, and the steering lines share the remaining part. In particular, the difference of pulling force between the two steering lines, $\Delta F$, depends roughly linearly on the difference of their length, according to the following approximate relationship:
\begin{equation}\label{E:line_difference}
\Delta F\doteq F_l-F_r\simeq\dfrac{h}{w_s^2}\rho\,A\,C_L\,E^2\,\left(1+\dfrac{1}{E^2}\right)^{\frac{3}{2}}\,W^2\delta,
\end{equation}
where $F_l,\,F_r$ are, respectively, the forces exerted on the left and right lines, $h$ is the wing's height, i.e. the distance between the line connecting the wing's tips and the center point of the leading edge (see Fig. \ref{F:wing_geom} for a graphical representation), and $\delta\doteq L_r-L_r$ is the difference between the length of the right steering line, $L_r$, and the left one, $L_l$. Equation \eqref{E:line_difference} is derived from a rotation equilibrium of the wing around the point where the center line force is applied, neglecting all forces except for the aerodynamic lift and drag and the lines' traction. The typical maximal value of $\delta$ during crosswind flight is given by:
\begin{equation}\label{E:max_delta}
\overline{\delta}=0.15\,w_s,
\end{equation}
which results in a roll angle $\psi$ of the wing of about 0.15$\,$rad (see e.g. \cite{FZMK13_arxiv}). From \eqref{E:line_difference} and \eqref{E:max_delta}, the maximal absolute value of $\Delta F$ can be approximately computed as:
\begin{equation}\label{E:max_deltaF}
\overline{\Delta F}=0.15\,\dfrac{h}{w_s}\rho\,A\,C_L\,E^2\,\left(1+\dfrac{1}{E^2}\right)^{\frac{3}{2}}\,W^2.
\end{equation}

\begin{figure}[!hbt]
\centering{
\includegraphics[width=8cm]{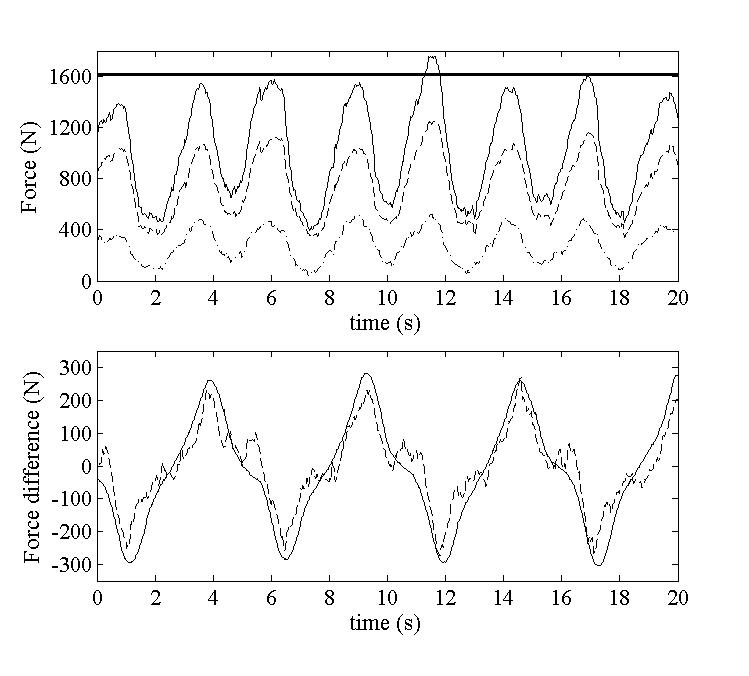}}
\caption{Experimental results. Upper plot: time courses of the total force (solid line) and of the force acting on the power line (dashed) and on the steering lines (dash-dotted). The thick solid line is the peak force value given by $\eqref{E:total_force}$. Lower plot: time courses of the difference $\Delta F$ of steering lines' forces, either measured  (dashed) or predicted by equation \eqref{E:line_difference} (solid). Wing parameters: $A=9\,$m$^2$, $C_L=0.8$, $E=5.6$, $w_s=2.7\,$m, $h=1.8\,$m. Wind speed $W=3.4\,$m/s.}\label{F:example_forces}
\end{figure}

As an example, a wing with $A=9\,$m$^2$, $C_L=0.8$, $E=5.6$, $w_s=2.7\,$m, $h=1.8\,$m would generate $\overline{F}\simeq1,600\,$N peak force during crosswind motion with $W=3.4\,$m/s wind speed \eqref{E:total_force}, the minimal force  would be $400\,$N and the period would be $T_L\simeq2.5\,$s \eqref{E:force_oscillations}-\eqref{E:path_length}. The power line would experience 55\% to 75\% of the load, and the remaining part would be experienced by the steering lines. The maximal difference of force between the steering lines, $\overline{\Delta F}$, would be about 320$\,$N \eqref{E:max_deltaF}. For a comparison, Fig. \ref{F:example_forces}
shows the experimental data measured on our small-scale prototype with the mentioned wing parameters and wind conditions. It can be noted that the provided guidelines match well with experimental evidence. A similar good matching was observed with different wings and different wind speeds.

Apart from the capability to withstand the forces exerted by the lines, the other requirements to be considered in the frame design are the availability of mounting points for the installation of the various components, easiness of assembly/modification, low weight for easy transportability and installation,  resistance to corrosion (particularly if the field tests are carried out in humid areas or in proximity of the sea). These aspects can be dealt with by a proper choice of materials and assembly solutions, see section \ref{S:design} for the details of the system that we realized at the University of California, Santa Barbara.

The linear motion systems have the function of converting the rotation of the prototype's electric motors into a linear movement, in order to issue a difference of the steering lines' length and to change the length of the power line relative to the steering lines. The former action gives rise to a steering deviation on the wing, while the latter is used to change its pitch angle and hence its aerodynamic coefficients. A typical solution employs lead screw mechanisms and a series of pulleys to redirect the lines. The main design requirements for these systems are therefore the capability to withstand the line forces, to actuate a sufficiently large change of linear position, and to reduce the loads that must be sustained by the electric motors through a suitable gear ratio. For the steering lines', a good value for the maximum length difference that can be actuated is $\pm\frac{1}{2}w_s$ (i.e. half the wingspan), while for the power line the range should be $\left[-\frac{1}{4}h,\,0\right]$, with 0 being the configuration used during crosswind flight and negative values giving a de-powering of the wing. The design loads for the LMSs can be computed with equations \eqref{E:total_force}-\eqref{E:max_deltaF}. As regards the pulleys used to redirect the lines, a minimum diameter of $30\,d_l$, where $d_l$ is the line diameter, should be used to limit the stresses on the lines due to excessive bending.

The last pulleys before the wing (``lead-out'' pulleys) must also allow a sufficiently wide range of line angles with respect to the frame. Ideally, the lead-out pulleys shall tolerate an angle of up to $\pm\pi\,$rad between the lines projected on the ground and the $X$ axis of the frame, with any elevation in the range $[0,\,\pi/2]\,$rad. However, if the GU is properly aligned with the absolute wind direction, lateral angles of $\pm\,\frac{\pi}{2}\,$rad are enough for take-off, crosswind flight and landing.

The details of the specific solutions that we used in our small-scale prototype for the LMSs and the pulleys are given in section \ref{S:design}.

\subsection{Electric motors and drives}\label{SS:motors}
In a typical design, one electric motor is used to actuate the steering command, and a second one to actuate the power/de-power (i.e. wing pitch) command. The chosen motors shall have a sufficiently high stall torque and rated speed to withstand the peak loads acting on the lines and to actuate the required inputs quickly enough. The involved force values for given wing and wind conditions can be computed as shown in section \ref{SS:frame}. For the maximal actuation speed, values of $\frac{1}{3}w_s/$s and $\frac{1}{4}h/$s for the steering lines' difference and the power line length, respectively, yield a fast enough actuation in all the usual wind conditions. As an example, for the 9-m$^2$ wing considered in section \ref{SS:frame}, the speed of the steering actuation  shall be around $1\,$m/s and that of the pitch actuation around $0.5\,$m/s. Clearly, the required motor speed and torque depend also on the design of the linear motion systems, i.e. the matching between the motors and the linear transmission has to be chosen in order to achieve the required speed and load capability. A typical choice is the use of DC brushed or brushless electric motors with incremental quadrature encoders for position feedback. Another important aspect is the energy consumption required for actuation: to this end, since the largest load are experienced at low or zero speed of the actuators, it is preferable to choose fast motors and high gear ratios (motor/linear motion) of the LMSs to keep the motor torques (hence currents) low. Finally, the chosen motors shall be sealed and rated for operation in the field.

As regards the motor drives, they shall be able to apply to the motors the reference currents issued by the real-time control hardware. The peak current given by the drives shall at least match the currents required by the motors to withstand the peak loads. Moreover, the drives shall provide the motor positions as feedback variables.  The low-level control loops for the motors' positions can be implemented on the real-time hardware or, as an alternative, drives with their own speed and/or position control loops can be used. Drives that can take standard AC current as input and can control DC brushed or brushless motors can be easily found and are relatively cheap. The mounting of the drives on the prototype shall take into account a proper ventilation and heat dissipation, as well as short and shielded signal cables between the drives and the real-time hardware and the drives and the motors' encoders, to minimize noise on the position feedback.

\subsection{Power supply}\label{SS:power}
The power supply system should be properly dimensioned to supply the required current and voltage to the motor drives, as well as to the real-time hardware and the sensors. Systems based on lead-acid batteries shall be preferred for their robustness and fast response to peak loads. A bank of 12$\,$V or 24$\,$V batteries connected in parallel can be used, together with an inverter providing pure sine AC current as output. The inverter power and peak current shall be large enough to supply both motors at peak loads, plus the sensors and the control hardware. The average current consumption can be derived on the basis of the loads that the motors must sustain during operation, divided by the motors' constants, plus the current drawn by the measurement and control instrumentation. The obtained current can then be translated from the motor side to the battery side. As an example, in a system with 120$\,$V, 60 Hz AC on the drives/control hardware side, the current drawn on the battery side (assuming 12$\,$V DC batteries) can be taken as 10 times larger than the one drawn on the drives side. The operation of the drives and of the measurement and control system in idle conditions, i.e. with zero current absorbed by the motors, can easily require 20$\,$A of continuous current on the battery side (i.e. 240 W and about 2$\,$A on the drives' side). On the basis of the computed currents, the battery bank can then be dimensioned in order to guarantee the desired number of hours of operation, e.g. 5 or 8 hours of crosswind flight in a single testing day, considering the nonlinear discharge time of the batteries as a function of the drawn current. The power supply system shall be designed in order to maximize safety and guarantee a proper ventilation of the inverter.

\subsection{Real-time control hardware}\label{SS:RThardware}
The control hardware comprises a real-time processor to execute all the necessary signal processing, estimation and control algorithms and data logging, as well as suitable interfaces to acquire and send digital and analog inputs/outputs. The interfaces should include several serial communication ports, which are often used by IMU (inertial measurement unit), GPS (global positioning system) and compass sensors, analog-to-digital converters used for example to acquire inputs from the Human-Machine Interface, and digital inputs for quadrature encoder signals and control switches. The computational power should be large enough to achieve 100$\,$Hz for the motor position control loops, and $50\,$Hz for the wing path controllers (see \cite{FZMK13_arxiv} for details of such control loops). Moreover, the amount of memory available for data logging shall be large enough to cover the desired test duration, denoted as $\overline{T}$ (in seconds). The required memory in Bytes can be easily estimated as:
\[
M=\overline{T}\,\sum\limits_{k=1}^{N_{T_s}}\frac{n_{s,k}\,B_{s,k}}{T_{s,k}},
\]
where $N_{T_s}$ is the total number of different sampling frequencies used in the control system and $n_{s,k},\,B_{s,k},\,T_{s,k}$ are, respectively, the number of signals, the number of Bytes for each signal and the sampling period (in seconds), pertaining to the $k-$th sampling frequency.
Systems specifically tailored for rapid prototyping, like National Instruments$^\circledR$ or xPC Target$^\circledR$ products, provide the best compromise between performance, robustness and versatility. The details of the design choices for our small-scale prototype are given in the section \ref{SS:power supply}.

\subsection{Human-Machine-Interface}\label{SS:HMI}
The HMI shall provide a human operator with the means to control the flight of the wing and to engage/disengage the automatic control system and, eventually, different types of semi-autonomous operating modes. Two commands shall be available, one to control the position of the LMSs for the steering deviation and one for the power/de-power setting of the center line. Such commands can be realized with an analog, 2-axis joystick where each axis is linked to the position of one actuator. The other inputs can be realized by means of simple switches. Moreover, events like excessive wind speed, low battery voltage or sensor failure can be also communicated to the human operator via buzzers or lights activated by the real-time hardware. All the mentioned devices can be found with different DC input and output voltage ranges, typically 0-5$\,$V and 0-10$\,$V, which shall match the analog and digital input/output channels available on the real-time hardware.

\subsection{Wings and lines}\label{SS:wings}
The wing is a crucial component of the prototype, as it influences a number of design specifications including the loads applied to the structure and the position range and speed of the motors and actuators, as shown in the previous sections. While the design of wings specifically tailored for airborne wind energy is a topic of current research (and indeed the designed wings can be tested with the prototype setup described in this paper), a functioning system can be achieved with commercially available, three-lined power kites. According to our experience, LEI (leading edge inflatable) kites provide the best behavior in terms of structural stability, aerodynamic efficiency and controllability. Kites with different sizes can be used to match with different wind conditions. To this end, a maximal resistance of the kite of about 250 N/m$^2$ can be considered and compared with the total force in crosswind conditions (given by \eqref{E:total_force}) in order to match the wing parameters with the wind conditions. When powered, LEI kites have lift coefficients ranging between 0.6 and 1 and efficiency values $\simeq5$. The de-power range of these kites is quite large, so that  a significant reduction of the pulling force can be obtained by shortening the center line.
 \begin{figure}[!hbt]
 \centerline{
 \includegraphics[width=7cm]{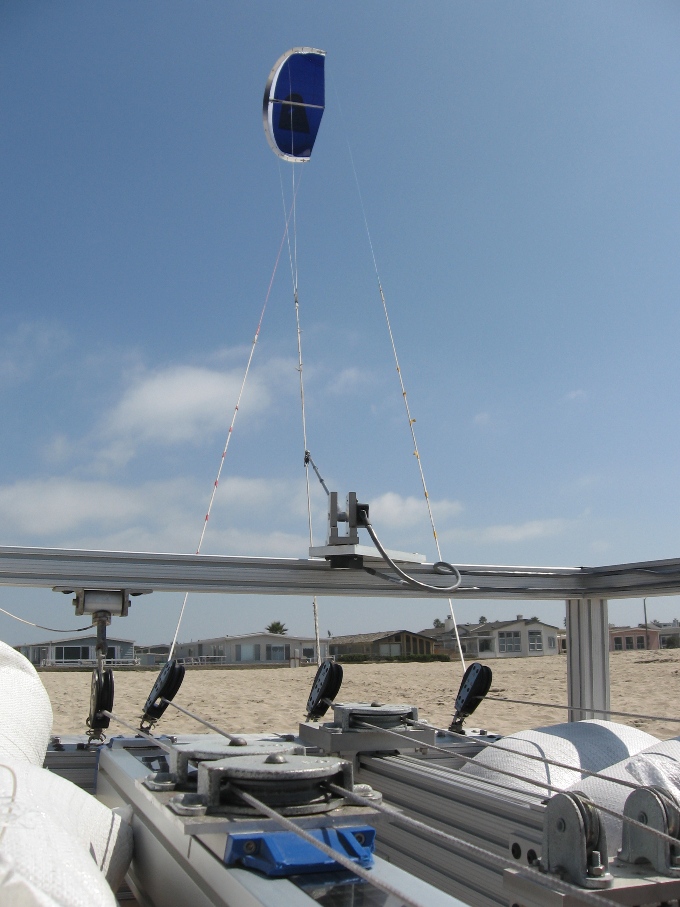}}
\caption{Small-scale prototype built at the University of California, Santa Barbara, to study the control of tethered wings for airborne wind energy. The linear actuation system for the steering deviation, a load cell, the line angle sensor, the LEI kite and its lines are visible in the picture.}\label{F:proto}
 \end{figure}

As regards the lines, ultra-high-molecular-weight polyethylene fibers are one of the best material for this application, the most common commercial product being Dyneema$^\circledR$.   Lines of various diameters with the related breaking loads are available for sale. The line diameter shall be chosen according to the expected peak loads, with a safety factor of 4-5. As an example, a line with 0.003$\,$m diameter has a minimum breaking load of around 1.1$\,10^4\,$N. Lines made with this material have a density of $970\,$kg/m$^3$. It has to be noted that the center line is split into two lines attached on the leading edge of the wing (see Fig. \ref{F:proto} for a picture), which share the total load. Pre-stretched lines shall be used, in order to avoid changes of line length during the first uses. The line length has to be chosen in order to match with the space available at the chosen testing sites and the local airspace regulations. Moreover, the line length has to be matched with the wing size and wind conditions, in order to avoid or limit line sagging effects. In our experience, line sagging is absent when LEI kites larger than 6$\,$m$^2$ are used, with wind speed $W>3\,$m/s and lines of 30$\,$m. The lines can be connected to the wing's bridles through standard knots.

\subsection{Sensors and radio communication}\label{SS:sensors}
\begin{figure}[!hbt]
 \centerline{
 \includegraphics[width=7cm]{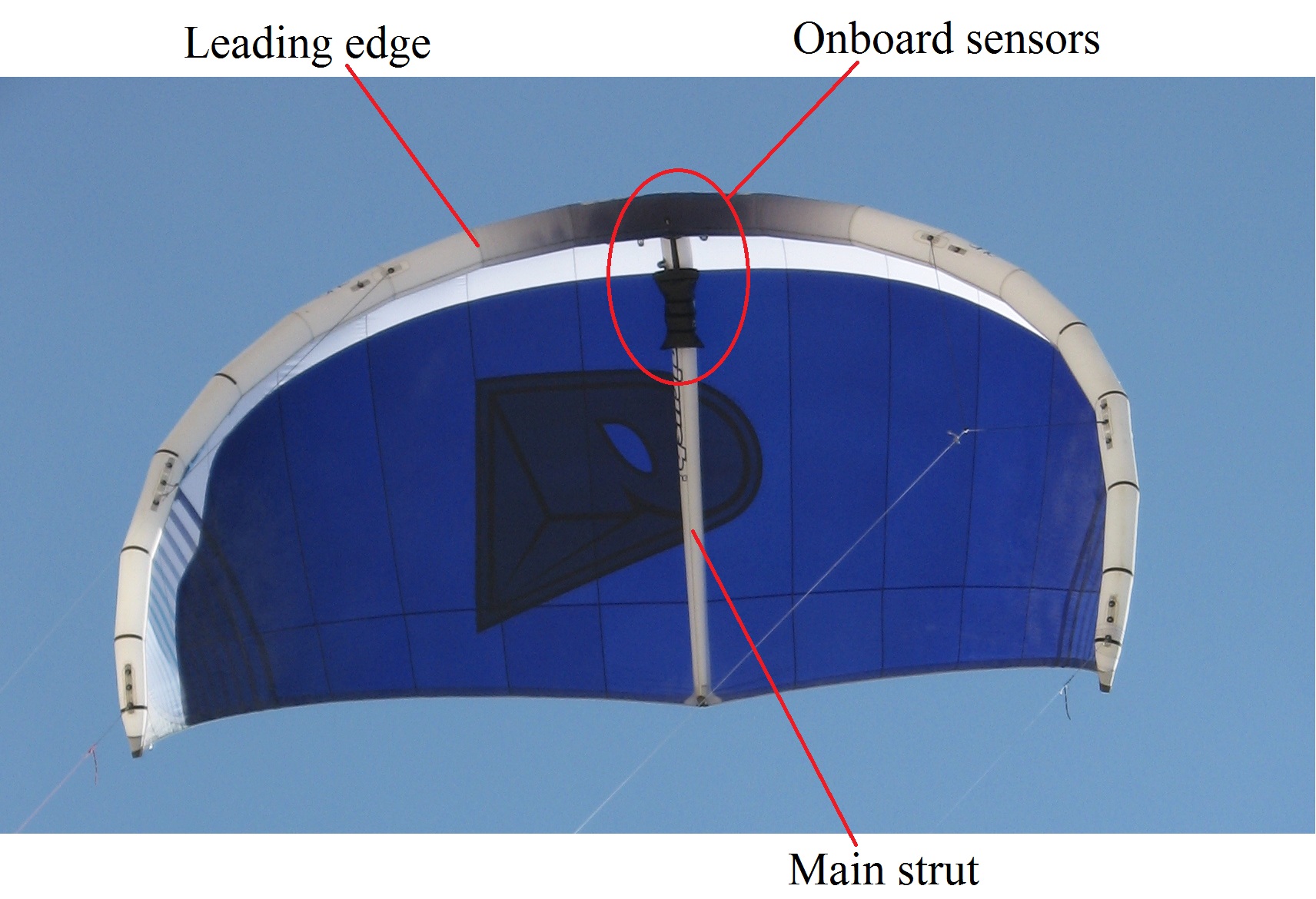}}
\caption{LEI wing used with the prototype. The black onboard sensors' package is fixed with velcro straps to the main strut, close to the leading edge. }\label{F:sensors}
 \end{figure}

The  ground and onboard sensors have to provide measurements of all the quantities of interest. For the ground unit, these include the azimuthal and elevation angles of the center line (and eventually of the two steering lines), the position of the GU (using a ground GPS), its orientation, the wind speed and direction, the load acting on each line, the position and current of the motors, the battery voltage, the motor currents. The onboard sensors shall provide the wing 3D accelerations and 3D angular velocities, position and linear velocity. IMUs with signal conditioning and filters are commercially available, or they can be developed ad-hoc for this application and tested on the prototype. The data acquired by the IMU can be sent to the ground control system via a radio link. Low transmission rates and radio frequencies  shall be used to improve the robustness of the communication (compatibly with the amount of data to be transmitted and the sampling frequency). A study on the sensors and sensor fusion aspects for the purpose of feedback control of the wing's flight is given in \cite{FHBK12_arxiv}, covering also the main specifications of the employed onboard sensors and of the line angle sensor. The latter has been developed ad-hoc for this application and we provide its detailed design in the next section, together with the specifications of the other chosen ground sensors. The onboard sensors have to be attached on the wing at a point that provides a stiff enough mounting. In LEI kites, a good solution is to attach the sensors' package to the main strut of the wing, just behind the leading edge (see Fig. \ref{F:sensors})

\section{Prototype design and basics of operation}\label{S:design}

In this section, we first provide the details of the specific design that we used for our prototype, focusing in particular on the frame, linear motion systems, and line angle sensor, as well as a cost breakdown. We then describe the basic maneuvers that can be undertaken with the prototype.

\subsection{Frame, linear motion systems and motors}\label{SS:Frame_actuators}
\begin{figure}[!hbt]
 \centerline{
 \includegraphics[width=8cm]{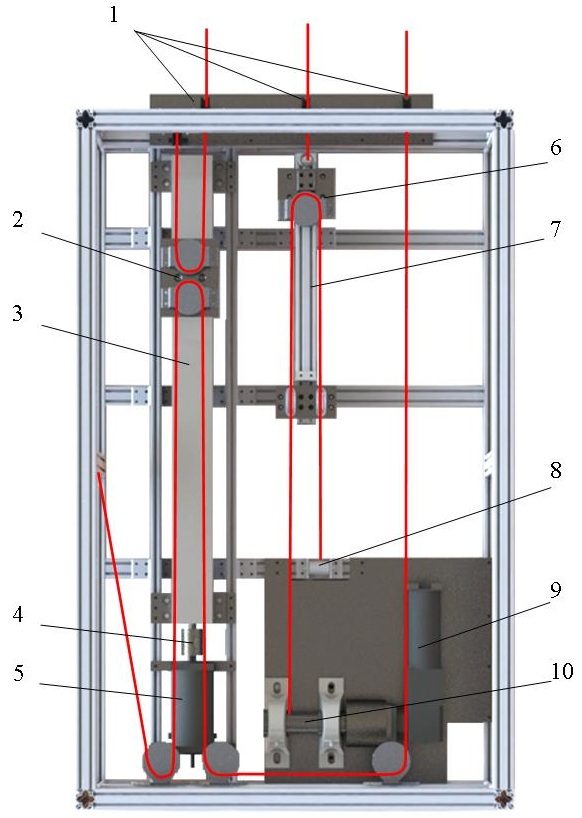}}
\caption{Rendering of the top view  of the designed frame. Thick red lines: paths of the three lines. 1: swaying lead-out pulleys; 2: carriage for the steering linear motion system (LMS); 3: steering LMS; 4:  LMS-motor coupling; 5: motor for the steering LMS; 6: carriage for the power/de-power input; 7: linear guide for the power/de-power input; 8: load cell for the center line; 9: right-angle gearmotor for the power/de-power input; 10: output shaft for the power/de-power input.}\label{F:frame_top}
 \end{figure}

\begin{figure*}[!hbt]
 \centerline{
 \includegraphics[width=18cm]{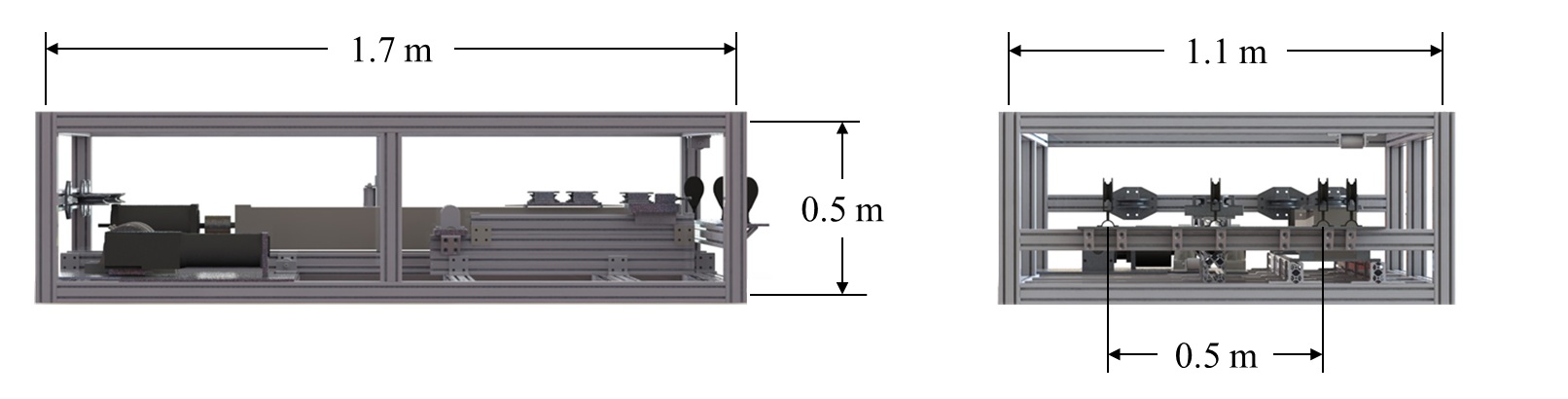}}
\caption{Rendering of the side and front views  of the designed frame. The outer dimensions are indicated, as well as the distance between the lead-out pulleys for the steering lines. }\label{F:frame_side}
 \end{figure*}

\begin{figure}[!hbt]
 \centerline{
 \includegraphics[width=8cm]{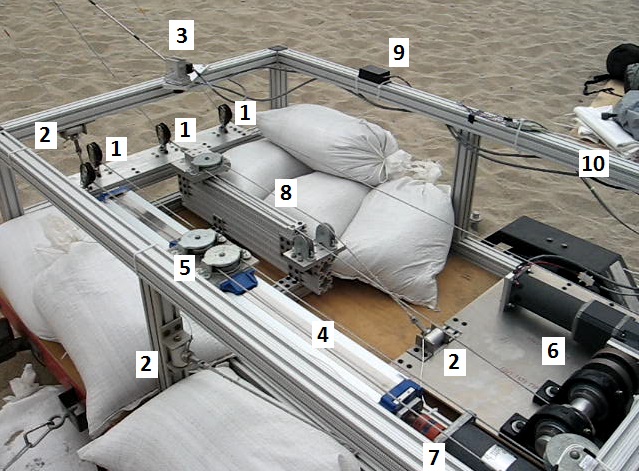}}
\caption{Overview of the constructed ground unit. 1: lead-out pulleys; 2: load cells; 3: line angle sensor; 4:  linear motion system for the steering input; 5: carriage for the steering LMS; 6: right-angle gearmotor for the power/de-power input; 7: motor for the steering input; 8: linear bearing; 9: ground compass; 10: mechanical frame.}\label{F:frame_pic}
 \end{figure}

We realized the frame using 80/20$^\circledR$ extruded aluminum T-slot beams, as they
allow for fine adjustments to be made to individual pulleys, thereby
increasing pulley alignment, while maintaining light weight, strength, corrosion resistance and ease of assembly. Figs. \ref{F:frame_top}-\ref{F:frame_side} show, respectively,
 the top view of the frame, with some key components highlighted, and the side and front views, with the main dimensions as well as the distance between the lead-out pulleys of the steering lines. The latter value has an important effect on the self-steering behavior of the wing during flight, see \cite{FZMK13_arxiv} for details. Fig. \ref{F:frame_pic} shows the final result installed on a small trailer.

 The LMS for the steering lines' difference is a lead screw mechanism manufactured by Thomson Industries, with a total travel of 1$\,$m. We designed and manufactured a carriage (see Fig. \ref{F:frame_top}, n. 2, and Fig. \ref{F:frame_pic}, n. 5) with two pulleys in order to translate the motion of the LMS into a difference of the lines' lengths.
With the chosen solution, a displacement of $x$ of the carriage from the central position is converted in a length difference of 4$\,x$ of the steering lines. We limited the commanded displacement of the LMS to 0.35$\,$m, hence obtaining a range of allowed steering inputs of $\pm1.4\,$m. The LMS for the steering input is connected to a brushed DC motor (Fig. \ref{F:frame_top}, n. 5, and Fig. \ref{F:frame_pic}, n. 7) manufactured by Groschopp Inc., with stall torque of 6.58$\,$Nm at 10$\,$A current, and 2400$\,$rpm maximum speed. The screw diameter of the LMS is 0.025$\,$m, and the transmission ratio is 0.01$\,$m per revolution of the motor. As regards the LMS for the center line, we used a simple linear guide (see Fig. \ref{F:frame_top}, n. 7, and Fig. \ref{F:frame_pic}, n. 8) and manufactured a carriage (Fig. \ref{F:frame_top}, n. 6) with a linear bearing and a pulley. The latter splits  the force exerted by the wing's center line between the output shaft (Fig. \ref{F:frame_top}, n. 10) of a right-angle gearmotor (Fig. \ref{F:frame_top}, n. 9, and Fig. \ref{F:frame_pic}, n. 6) and a load cell (Fig. \ref{F:frame_top}, n. 8, and Fig. \ref{F:frame_pic}, n. 2), in order to halve the torque applied on the gearmotor shaft (whose radius is $0.012\,$m) and, at the same time, to allow for a measurement of the force on the center line.  The right-angle gearmotor has a stall torque (at the output of the gearbox) of 63$\,$Nm at 10$\,$A current, and maximum speed of 240$\,$rpm. The rotation of the gearmotor gives rise to a displacement of the carriage, hence changing the length of the center line. In particular, this system can shorten the center line by up to 0.5$\,$m with respect to the initial setting.\\
We built the lead-outs (shown in Fig. \ref{F:frame_pic}, n. 1) using pulleys manufactured by Harken, which we assembled with coil springs (visible also in the movie \cite{Wing_movie}).
\begin{figure}[!hbt]
 \centerline{
 \includegraphics[width=8cm]{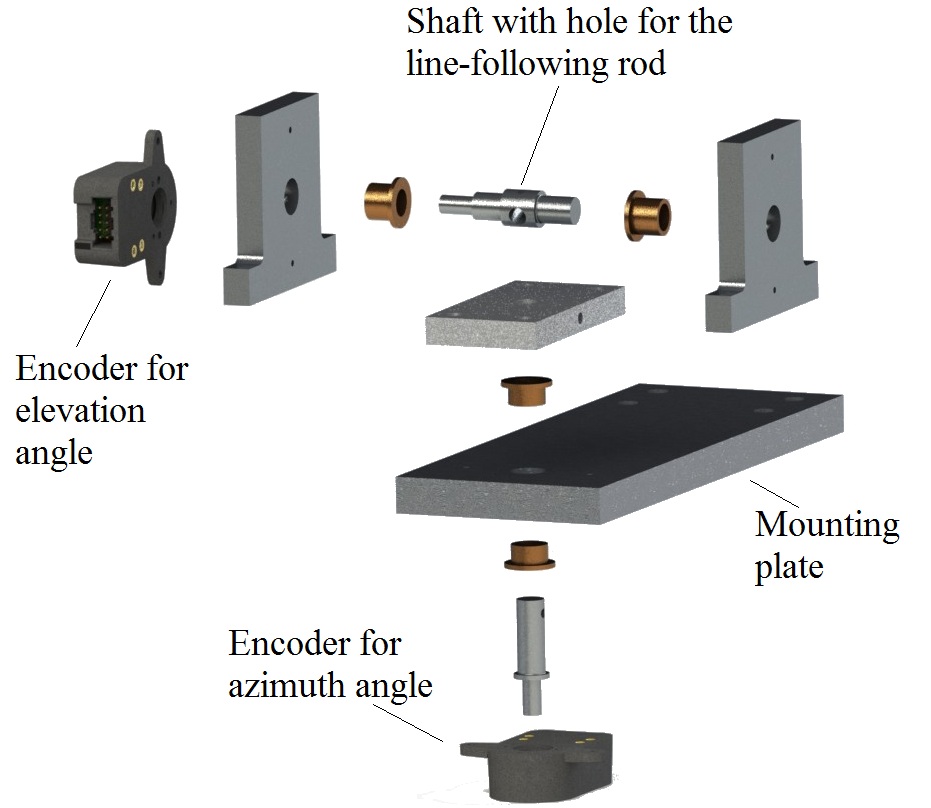}}
\caption{Schematic of the line angle sensor.}\label{F:line_angle}
 \end{figure}
 \begin{figure}[!hbt]
 \centerline{
 \includegraphics[width=6cm]{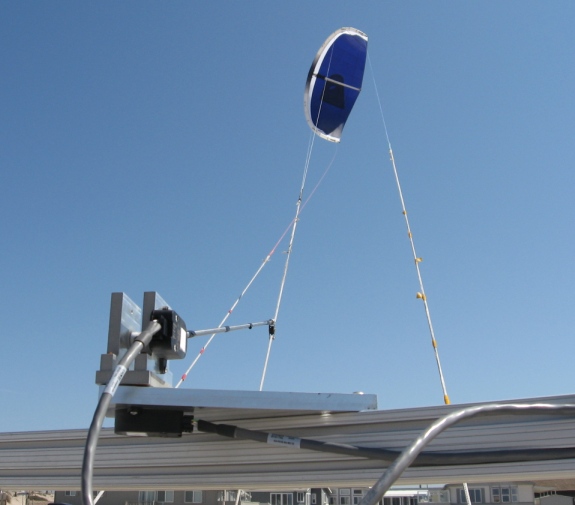}}
\caption{Line angle sensor.}\label{F:line_angle_pic}
 \end{figure}
\subsection{Line angle sensor}\label{SS:line_angle}
The line angle sensor is crucial to obtain accurate measurements of the wing's position and to estimate its velocity. We built such a sensor by using two shafts: the first one, for the azimuth angle, is allowed to rotate around an axis perpendicular to the ground; the second one, for the elevation angle, is fixed to the first (hence it rotates around the vertical axis) and is allowed to rotate around an axis parallel to the ground. A steel rod is fixed to the second axis and it is attached to the center line. The movement of the line drives the steel rod, which in turn causes the two shafts to rotate. Two encoders measure the shafts' rotations to provide the azimuthal and elevation angles of the wing. The details of the functioning of this sensor are given in \cite{FHBK12_arxiv}, together with the related sensor fusion algorithms. A detailed scheme of the device is shown in Fig. \ref{F:line_angle}, while a picture is shown in Fig. \ref{F:line_angle_pic}. The employed encoders are differential optical quadrature encoders manufactured by US Digital, with 400 counts per revolution.

\subsection{Power supply, drives, control hardware/software  and human-machine interface, sensors, wings and tethers}\label{SS:power supply}

\begin{figure}[!hbt]
\centerline{
\includegraphics[width=6cm]{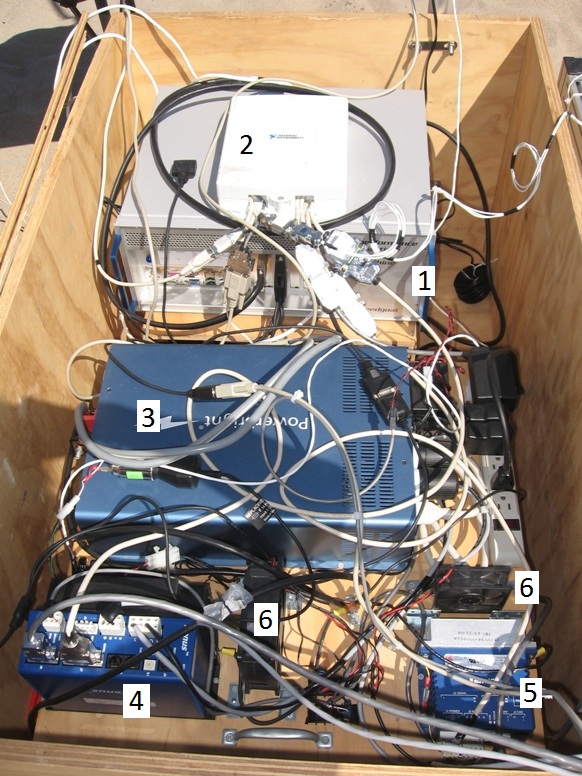}}
\caption{Electric and electronic components of the prototype. 1: real-time machine; 2: analog and digital signals interface; 3: DC-AC inverter; 4: drive for the steering motor; 5: drive for the center line motor; 6: cooling fans. 16 lead-acid batteries connected in parallel are placed below the instrumentation shown in the picture.}\label{F:battery_RTM_n}
 \end{figure}

We used 16 lead acid batteries in parallel, each one with 20$\,$Ah capacity at 1$\,$A discharge current, to build a 12$\,$V DC power supply, which we connected to a pure sine inverter with 1,500$\,$W nominal power (see Fig. \ref{F:battery_RTM_n}, n. 3). The drives we used for the steering motor and for the center line motors are, respectively, a Xenus$^\circledR$ XTL-230-36 and a Xenus Micro$^\circledR$ XSJ-230-10 manufactured by Copley Controls (Fig. \ref{F:battery_RTM_n}, n. 4 and 5).
The real-time machine (Fig. \ref{F:battery_RTM_n}, n. 1) is manufactured by SpeedGoat$^\circledR$ and programmed using the xPC Target$^\circledR$ toolbox of Matlab$^\circledR$. The employed data acquisition interface is composed by a National Instruments$^\circledR$ PCI-6221 DAQ card and by a Quatech QSC-100-D9 with 4 RS232 ports. For the HMI, we used a 2-axis analog joystick (M21C051P by CH Products) and a switch to change from manual flight mode to automatic flight mode.

As regards the employed sensors, the details of the IMU and of the radio link are given in \cite{FHBK12_arxiv}. To measure the line forces, we used three load cells ELPF-T3 E-500L/10F/AMP by Measurement Specialties, installed at the attachment points of the lines on the frame (Fig. \ref{F:frame_pic}, n. 2). To measure the GU orientation, we used a OS5000 digital compass by Ocean Server Technology, Inc., while for the GU position we employed a GPS-18x by Garmin.

In our tests, we used three LEI kites with different sizes: a 6$\,$m$^2$, a 9$\,$m$^2$ and a 12$\,$m$^2$ Airush One$^\circledR$ 2012, whose parameters are reported in \cite{FZMK13_arxiv}. Finally, we used 27-m-long standard kite lines with a breaking load of 2600$\,$N each. We joined the flying lines with the Dyneema lines on the GU with standard lark's head knots used in kite surfing. The total line length is 30$\,$m.

\subsection{Cost breakdown}\label{SS:cost}
Table \ref{T:costs} shows the costs of the various components of the prototype, as well as the corresponding shares in the total cost. It can be noted that the sensors and control hardware account for almost half of the total cost. However, this cost would remain the same also for systems of larger sizes, while the other costs would increase.

\begin{table}
\centering \caption{Cost breakdown for the prototype built at the University of California, Santa Barbara}
{%
\begin{tabular}{|l|r|r|}
\hline\highertop
Component  & Cost (\$) & Share (\%) \\\hline\hline\highertop
\textbf{Sensors and control hardware} &\textbf{13,430.00} & \textbf{44.6}\\ \hline\hline\highertop
Inertial measurement unit& 4,430.00
& 14.7\\\hline\highertop
Real-time machine&3,800.00
& 12.6\\\hline\highertop
Load cells (3) & 1,400.00
& 4.6\\\hline\highertop
Analog/Digital  board and interface&1,240.00
& 4.2\\\hline\highertop
Radio modems& 900.00
& 3.1\\\hline\highertop
Line angle sensor & 400.00
& 1.3\\\hline\highertop
Ground compass& 330.00
& 1.1\\\hline\highertop
Ground anemometer&310.00
& 1.0\\\hline\highertop
Serial communication board& 220.00
& 0.7\\\hline\highertop
Joystick& 160.00
& 0.5\\\hline\highertop
Onboard sensors batteries and charger& 160.00
& 0.5\\\hline\highertop
Ground GPS& 80.00
& 0.3\\\hline\hline\highertop
\textbf{Mechanics and linear motion systems}& \textbf{7,400.00}&\textbf{24.6}\\ \hline\hline\highertop
Steering linear motion system & 4,100.00
& 13.6\\\hline\highertop
Framing & 2,000.00
& 6.7\\\hline\highertop
Manufacturing costs for ad-hoc parts& 1,300.00
& 4.3\\\hline\hline\highertop
\textbf{Miscellaneous}&\textbf{4,585.00} & \textbf{15.2}\\ \hline\hline\highertop
Consumables (wiring, connectors, etc.)& 2,500.00
& 8.3\\\hline\highertop
LEI kites (3)& 1,785.00
& 5.9\\\hline\highertop
Kite lines (3 sets) & 300.00
& 1.0\\\hline\hline\highertop
\textbf{Drives and motors} & \textbf{3,370.00}&\textbf{11.2} \\\hline\hline\highertop
Drive for the steering motor& 1,150.00
& 3.8\\\hline\highertop
Drive for the power/de-power gearmotor& 820.00
& 2.7\\\hline\highertop
Power/de-power gearmotor & 720.00
& 2.4\\\hline\highertop
Steering motor & 680.00
&  2.3\\\hline\hline\highertop
\textbf{Power supply}& \textbf{1,340.00}& \textbf{4.4}\\ \hline\hline\highertop
Batteries (16) &870.00
& 2.9\\\hline\highertop
Inverter &470.00
& 1.5\\\hline\hline\highertop
\textbf{Total}& \textbf{30,125.00}& \textbf{100.0}\\\hline
\end{tabular}\label{T:costs} }
\end{table}

\subsection{Basic flight maneuvers}\label{SS:maneuvers}
We describe next the three main maneuvers that need to be performed in order to carry out experimental tests with the prototype system. All three maneuvers can be seen in a movie available online \cite{Wing_movie}.

\begin{figure}[!hbt]
\centerline{
\includegraphics[width=8cm]{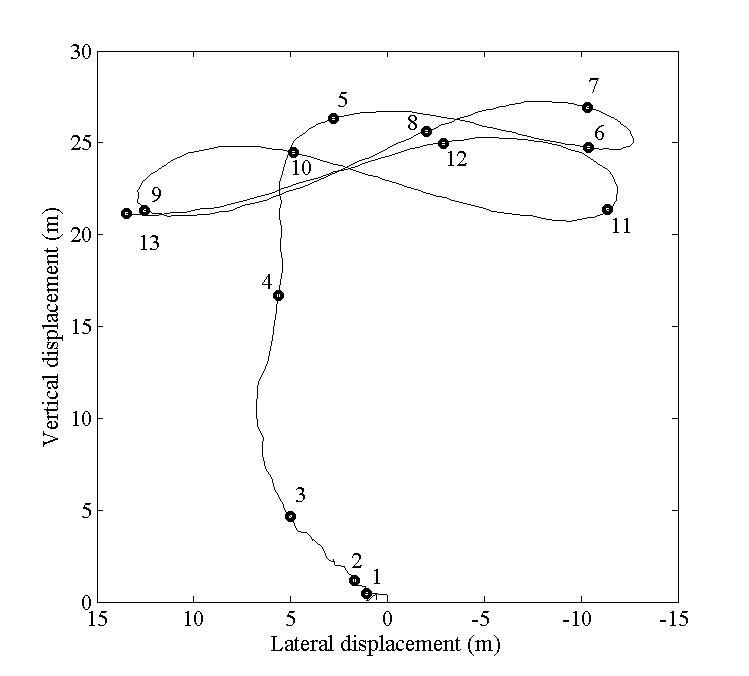}}
\caption{Experimental results. Trajectory of the wing  during take-off and crosswind flight, projected on a vertical plane perpendicular to the wind, as seen from the ground unit. The numbered points correspond to the time instants highlighted in Fig. \ref{F:take-off2}. Employed wing: 12-m$^2$.}\label{F:take-off}
 \end{figure}

\begin{figure}[!hbt]
\centerline{
\includegraphics[width=8cm]{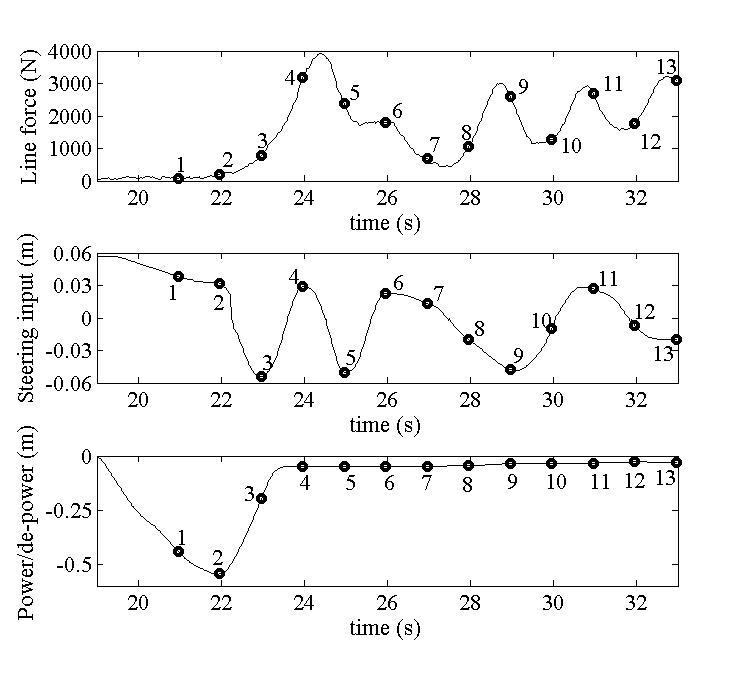}}
\caption{Experimental results. Time courses of the total line force, steering input and center line power/de-power input during take-off and crosswind flight. The steering input is in meters of displacement of the LMS carriage, the actual line length difference (right minus left) is four times larger than the plotted values. A negative position of the power/de-power input means a shorter center line (i.e. a de-powered configuration). The numbered points correspond to those highlighted in Fig. \ref{F:take-off}. Employed wing: 12-m$^2$.}\label{F:take-off2}
 \end{figure}

\textbf{Take-off}. If low or medium wind is present, the most reliable and quick take-off maneuver starts with the wing in downwind position, the lines aligned with the wind direction and the leading edge facing the sky. In this situation, the angle of attack of the wing is quite large, hence the lift force is generally low. However depending on the employed bridle setting and on the wind speed, a LEI wing might self-launch even in this conditions. Otherwise, the center line can be pulled to decrease the wing's angle of attack and point the leading edge into the wind in order to start the take-off. The wing's acceleration is quite large at take-off, and the wing's speed and force rapidly increase. As the wing approaches the vertical position, forces and speed decrease. During take-off, some steering corrections are needed to ensure that the wing gets airborne with a vertical trajectory. Figs. \ref{F:take-off}-\ref{F:take-off2} show experimental data related to the wing's trajectory,  line forces and inputs during the described take-off maneuver.\\
In the presence of strong wind, it is preferable to carry out the take-off by starting from a lateral position with respect to the wind direction, and gradually ``climbing'' the edge of the so-called ``wind window'',
\begin{figure}[!hbt]
\centerline{
\includegraphics[width=8cm]{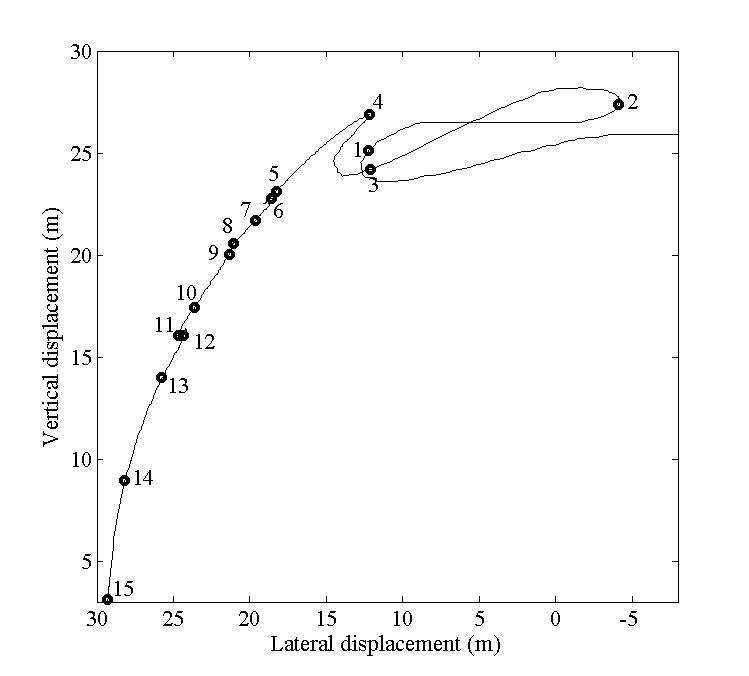}}
\caption{Experimental results. Trajectory of the wing  during landing, projected on a vertical plane perpendicular to the wind, as seen from the ground unit. The numbered points correspond to the time instants highlighted in Fig. \ref{F:landing2}. Employed wing: 12-m$^2$.}\label{F:landing}
 \end{figure}
 \begin{figure}[!hbt]
\centerline{
\includegraphics[width=8cm]{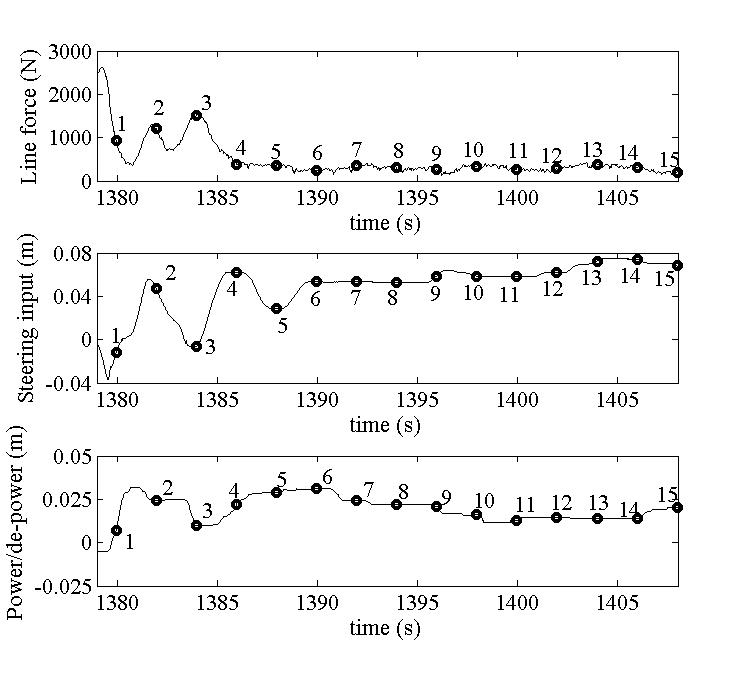}}
\caption{Experimental results. Time courses of the total line force, steering input and center line power/de-power input during take-off and crosswind flight. The steering input is in meters of displacement of the LMS carriage, the actual line length difference (right minus left) is four times larger than the plotted values. A negative position of the power/de-power input means a shorter center line (i.e. a de-powered configuration). The numbered points correspond to those highlighted in Fig. \ref{F:landing}. Employed wing: 12-m$^2$.}\label{F:landing2}
 \end{figure}
i.e. the quarter sphere defined by the ground, by a vertical plane perpendicular to the wind, and by the spherical surface that can be spanned by the wing's lines (see \cite{FZMK13_arxiv} for a formal definition). Such a maneuver is more difficult than the previous one, as it involves the control of the wing in almost-stationary conditions, with consequent low aerodynamical forces and stronger effects of gravity and external disturbances on the flight trajectory.

\textbf{Crosswind flight}. In this operating condition, the wing is controlled to fly fast along figure-eight paths, roughly perpendicular to the wind flow. The generated forces can be reduced, if needed, by moving the flown trajectories towards the top of the wind window. A detailed description of an automatic control system for crosswind flight is reported in \cite{FZMK13_arxiv}, together with extensive experimental data. A human operator is able to obtain quite good flight trajectories using the HMI, with relatively little training, by considering that a constant input gives rise to a constant clockwise (or counter-clockwise, depending on which line has been shortened, see \cite{FZMK13_arxiv}) turning speed of the wing. The data pertaining to two figure-eight trajectories carried out by a human operator right after take-off are shown in Figs. \ref{F:take-off}-\ref{F:take-off2}.

\textbf{Landing}. The landing maneuver consists in bringing the wing to the edge of the wind window, in a lateral position with respect to the wind velocity vector, and then making it gradually climb down  by applying a slight pull of the line closer to the ground. The line force is very low in this conditions, and the steering input has to be kept constant in order to achieve a constant descent speed of the wing. Experimental data related to the landing phase are shown in Figs. \ref{F:landing}-\ref{F:landing2}.

\section{Conclusions and future developments}\label{S:conclusion}
We presented the main design guidelines and the details of a specific solution for a small-scale prototype to carry out research in the field of airborne wind energy. The presented system allows to perform significant tests for a number of aspects, including aerodynamics and wing design, line and wing fatigue, sensors, controls, system modeling, actuation systems. The described design can be realized with low costs and relatively low effort by researchers approaching this field, hence contributing to reduce the initial difficulties in realizing an experimental test-bed. The described basic maneuvers can be used to operate the system. Finally, the prototype can be easily moved due to its lightweight, and it can be used without particular permissions in most places, if lines up to 50-60$\,$m of length are used.

The low cost of the prototype is achieved by not including power generation capabilities. However, the presented system's layout can be upgraded to add motor/generators and mechanics able to achieve line reeling. In the simplest configuration, a single motor/generator (e.g. a synchronous servomotor) can be connected via a gearbox to a drum, where the three wing's lines can be coiled. The motor/generator must be able to develop enough torque to control the line reel-out during the traction phases of the energy generation cycle. The guidelines provided here can be applied to dimension the motor/generator and the other components. Moreover, a line spooling system has to be added, using small motors and linear motion systems, to ensure that the lines are wound correctly on the drum. To provide an order of magnitude, a motor/generator with 15 kW rated power and around 70 Nm stall torque can be used together with LEI kites with sizes from 6 to 15$\,$m$^2$, like those used in our experimental setup. Line reeling can be also used to modulate the wing's force, which drops as the line reel-out speed increases (see e.g. \cite{FaMP11}). Examples of systems with generation capabilities can be found in \cite{CaFM09c,FaMP09,Craighton12}. The voltage of the DC power supply system of a prototype with energy generation would be between 300 and 600 V, in order to match with the bus voltage of the motor/generators drives. The total cost of such a prototype would be around 80,000$\,$\$. The next step in this research will be to design and build such a  prototype, and to develop an automatic control system able to carry out consistently the desired energy generation cycle, in order to assess experimentally the matching between the generated energy and the one predicted by numerical and theoretical analyses.

\end{document}